\documentclass[manuscript]{aastex6bis}

\usepackage{graphics}
\usepackage{lineno}

\begin{document}



\title{Ram pressure feeding super-massive black holes}


\author{Bianca M. Poggianti$^1$,
Yara Jaffe'$^2$, Alessia
Moretti$^1$, Marco Gullieuszik$^1$, Mario Radovich$^1$, Stephanie
Tonnesen$^3$, Jacopo
Fritz$^4$, Daniela Bettoni$^1$, Benedetta
Vulcani$^{5,1}$, Giovanni Fasano$^1$, Callum 
Bellhouse$^{6,2}$, George Hau$^2$, Alessandro Omizzolo$^7$}
\affil{$^1$INAF-Astronomical Observatory of Padova,   
vicolo dell'Osservatorio 5,   
35122 Padova, Italy, 
$^2$European Southern Observatory, Alonso de Cordova 3107,
Vitacura, Casilla 19001, Santiago de Chile, Chile, $^3$ Carnegie
Observatory, 813 Santa Barbara Street, Pasadena, CA91101, USA,
$^4$Instituto de Radioastronomia y Astrofisica, UNAM, Campus 
 Morelia, A.P. 3-72, C.P. 58089, Mexico, 
$^5$School of Physics, The University of Melbourne, Swanston St \&
 Tin Alley Parkville, VIC 3010, Australia, $^6$University of Birmingham School of Physics and Astronomy,
Edgbaston, Birmingham, England, $^7$Vatican Observatory, Vatican City State }






\section{First paragraph}

{\bf
When supermassive black holes at the center of galaxies accrete matter
(usually gas), they give rise to highly energetic phenomena named
Active Galactic Nuclei (AGN)$^{1,2}$. A number of physical processes
have been proposed to account for the funneling of gas towards the
galaxy centers to feed the AGN. There are also several physical processes
that can remove (“strip”) gas from a galaxy$^{3}$, and one of them is ram
pressure stripping in galaxy clusters due to the hot and dense gas 
filling the space between galaxies$^{4}$.
We report the discovery of a strong connection between severe ram
pressure stripping and the presence of AGN activity. 
Searching  in galaxy clusters at low redshift, we have selected
the most extreme examples of “jellyfish galaxies”, which are galaxies
with long “tentacles” of material extending for dozens of
kpc beyond the galaxy disk$^{5,6}$.
Using the MUSE spectrograph on the ESO Very Large Telescope, we find 
that 6 out of the 7 galaxies of this sample host a central AGN,
and two of them also have galactic-scale AGN ionization cones. 
The high incidence of AGN among the most
striking jellyfishes may be due to ram pressure causing gas to flow 
towards the center and triggering the AGN activity, or to an enhancement
of the stripping caused by AGN energy injection, or both. Our analysis
of the galaxy position and velocity relative to the cluster
strongly supports the first hypothesis, and puts forward ram pressure
as another, yet unforeseen, possible mechanism for feeding the central 
supermassive black hole with gas. 
}

\section{Main text with figures}

Black holes of different sizes are very common in the Universe. 
It is now well established that most, if not all, galaxies host at their center 
a supermassive black hole of a few million to a few billion 
solar masses$^{7,8}$. 
When a black hole accretes matter, 
it converts 
the gravitational energy of the accreted matter into mechanical and 
electromagnetic energy, giving rise to some of the most energetic 
astrophysical phenomena: Active Galactic 
Nuclei (AGN). 

One of the central questions regarding AGN is why if supermassive black 
holes are present in most galaxies, only a small fraction of these are 
AGN, i.e. why only a few of them are accreting matter.  It is believed 
that the black hole growth must be episodic, last typically 
$10^7-10^8$yr and that it must be related to a mechanism that drives 
efficiently gas to the galaxy center. 
Major mergers 
of two galaxies are among the best candidates for the most luminous 
AGN$^{9}$, while galaxy internal instabilities (e.g. 
driven by galaxy bars) or fast tidal encounters between galaxies might 
account for less luminous systems$^{10,11}$.

A prerequisite for AGN activity is therefore the availability of gas 
in the galaxy disk to feed the black hole. In the current cosmological 
paradigm, the interstellar medium present in the galaxy disk gets 
consumed by the formation of new stars but is continuously 
replenished by the cooling of hot gas present in the galaxy dark 
matter halo$^{12}$. 

However, there are several physical processes concurring to remove gas from 
galaxies especially in dense environments such as galaxy clusters 
and groups$^{3}$.
Ram pressure stripping due to the pressure exerted by the 
intergalactic medium on the galaxy interstellar medium is considered 
the most efficient of such processes$^{4}$. 
The galaxy loses its gas because the ram pressure overcomes the 
local binding energy, and in those regions of the galaxy where gas is 
removed, the formation of new stars is inhibited. 
However, before quenching the star formation, ram pressure can produce 
an enhancement of the star formation rate, as thermal instabilities 
and turbulent motions provoke the collapse of molecular
clouds$^{13,14}$.

The most spectacular examples of galaxies undergoing gas stripping by 
ram pressure are the so called ``jellyfish galaxies'', named this way 
because they have ``tentacles'' (tails) of gas and newly born stars 
that make them resemble the animal jellyfishes$^{5,6}$.

In this work, we show that there is a close link between strong ram pressure and AGN 
activity in jellyfish galaxies, establishing for the first time a 
probable causal connection between the two phenomena.
Our findings are based on GASP (GAs Stripping Phenomena in galaxies
with MUSE$^{15}$, \url{http://web.oapd.inaf.it/gasp}), 

which is an ESO Large 
Program aimed at studying where, how and why gas can get removed from galaxies. 
GASP studies 94 z=0.04-0.07 galaxies in clusters, groups and the field 
selected from optical images to have unilateral debris and asymmetric morphologies 
suggestive of gas-only removal mechanisms.
Spatially resolved gas and stellar 
kinematics and physical properties are obtained 
with the MUSE spectrograph on the Very Large Telescope. 

For the present work, we have selected all the cluster 
jellyfishes observed so far by GASP which have striking tails/tentacles 
of ionized gas, as seen by MUSE in the $\rm H\alpha$ line in emission at 6563
angstrom (\AA). We 
have selected those galaxies whose $\rm H\alpha$ tentacles are at least as long as 
the galaxy stellar disk diameter (see Extended Data Table~1). 
These are all massive galaxies, with stellar masses between $\sim 4 \times 10^{10}$ and $\sim 3 
\times 10^{11} M_{\odot}$. 

The $\rm H\alpha$ velocity maps of the 7 galaxies selected 
are shown in Figs.~1 and ~2 ((b) and (c) panels) and contrasted with the 
corresponding stellar velocity maps ((a) panels). 
The figure illustrates the 
long extraplanar ionized gas tentacles 
extending out to between $\sim 20$ and 
$\sim 100$ kpc. 

In contrast, the stellar velocity field is regular and shows that the 
stellar kinematics is undisturbed by the force acting on the 
gas. The comparison between the gaseous and stellar morphologies and 
velocity maps shows that these galaxies are undergoing a gas-only removal 
mechanism due to the impact of the intracluster medium (ICM) such as ram 
pressure stripping. Ram pressure calculations supporting this 
hypothesis for some of these galaxies are presented in the individual galaxy studies$^{15,16,17}$.

\begin{figure*}
\centerline{\includegraphics[scale=1.0]{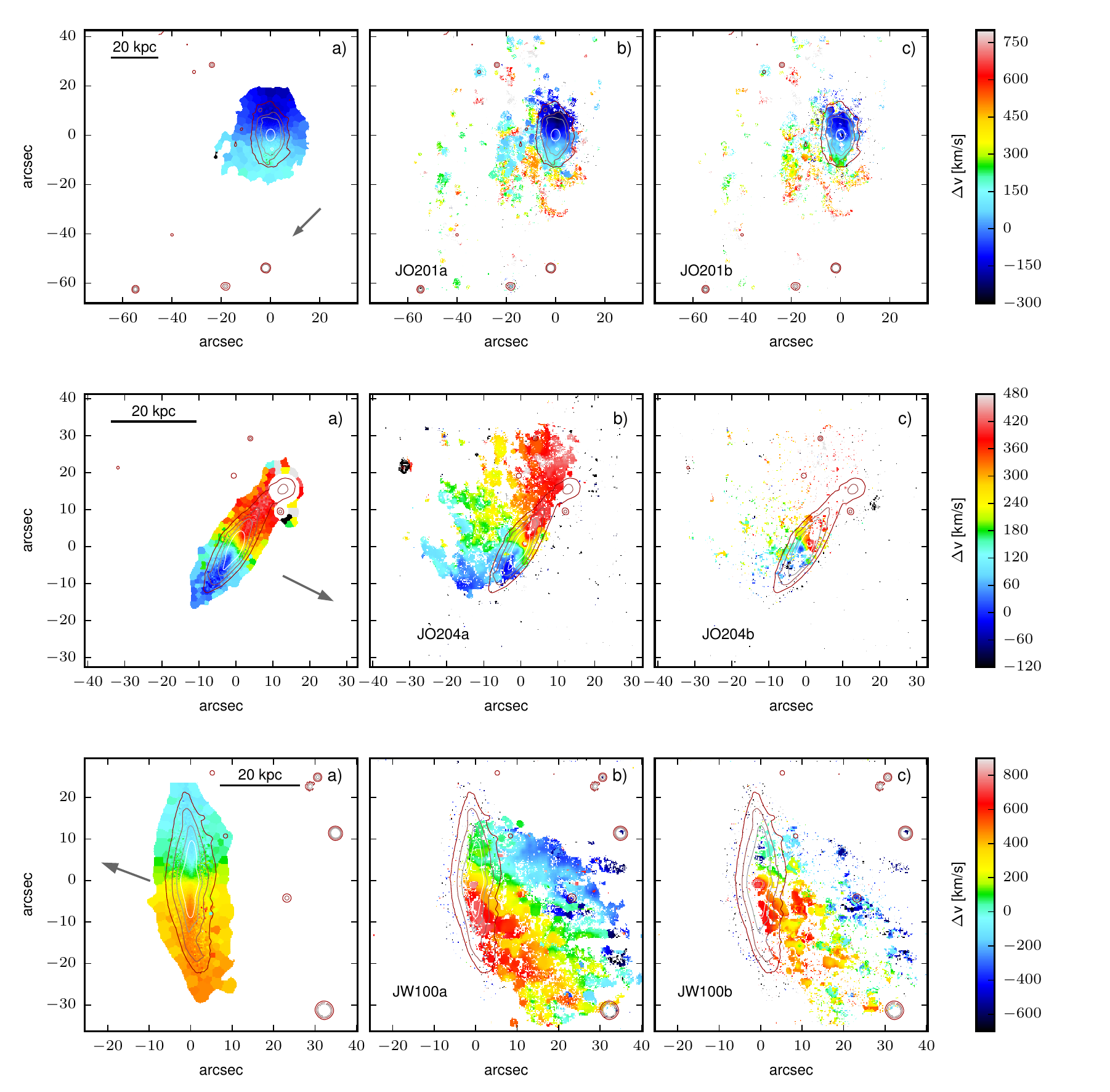}}
\caption{}
\end{figure*}

\begin{figure*}
\centerline{\includegraphics[scale=1.0]{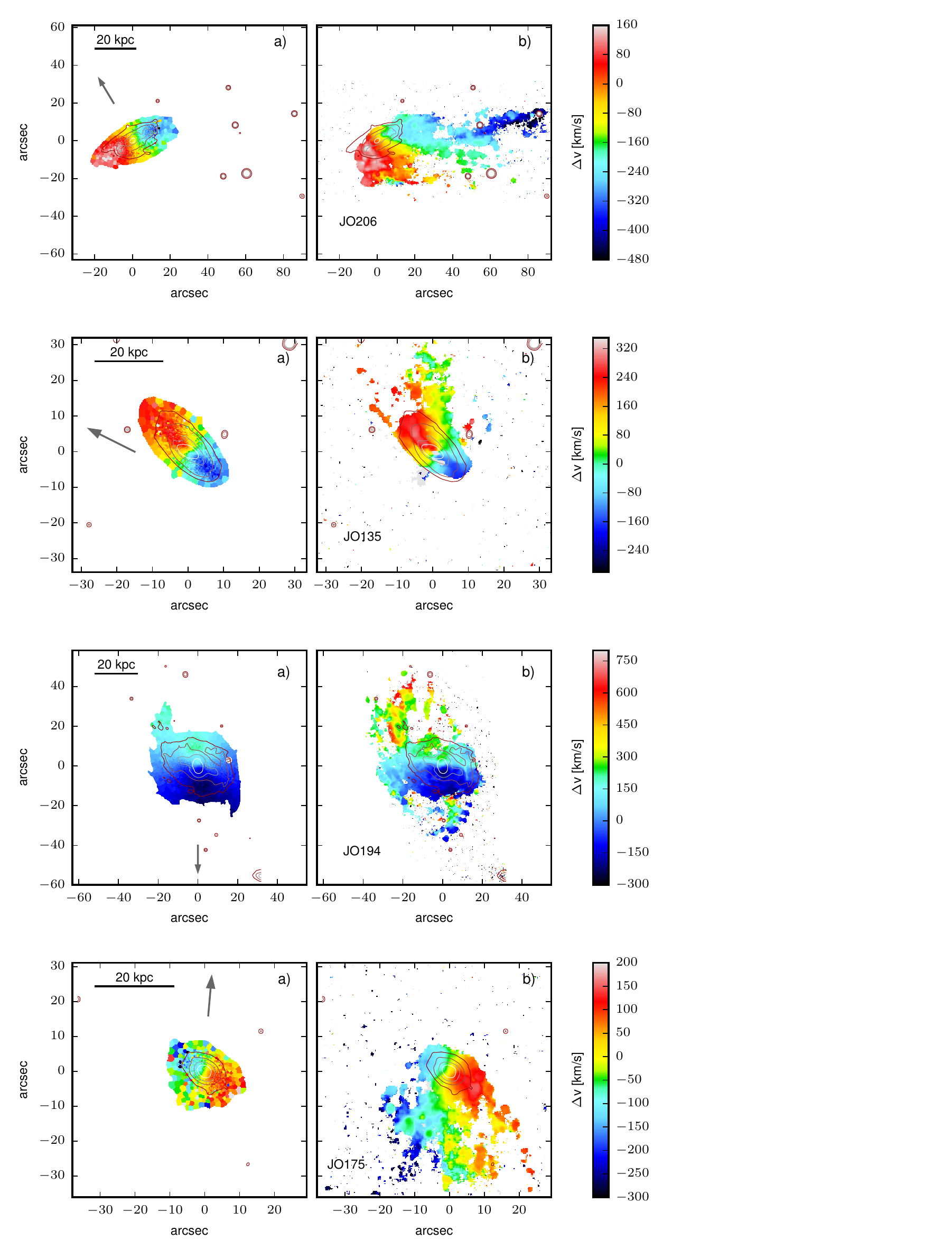}}
\caption{}
\end{figure*}

\begin{figure*}
\centerline{\includegraphics[scale=0.8]{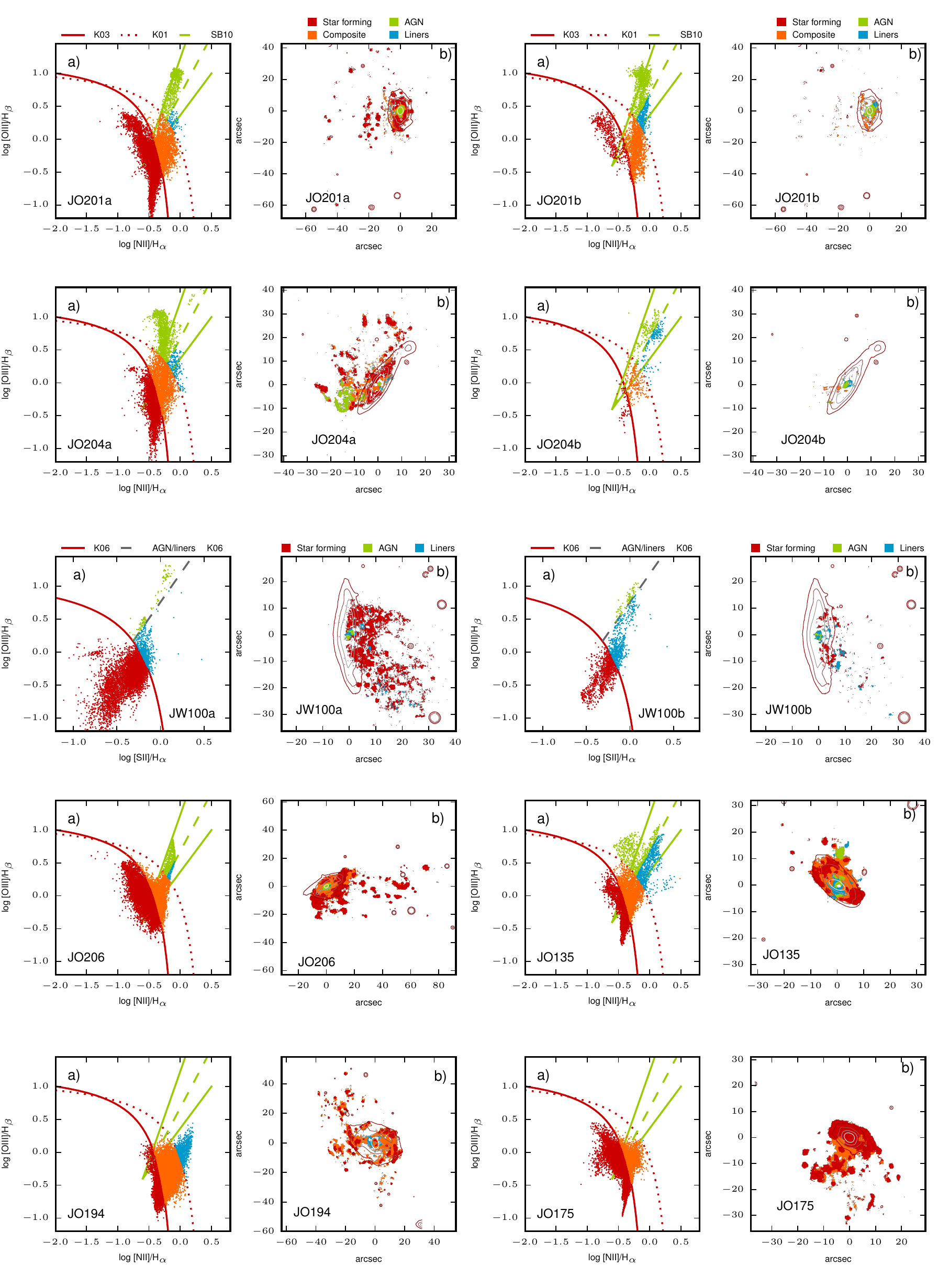}}
\caption{}
\end{figure*}

\begin{figure*}
\centerline{\includegraphics[scale=1.0]{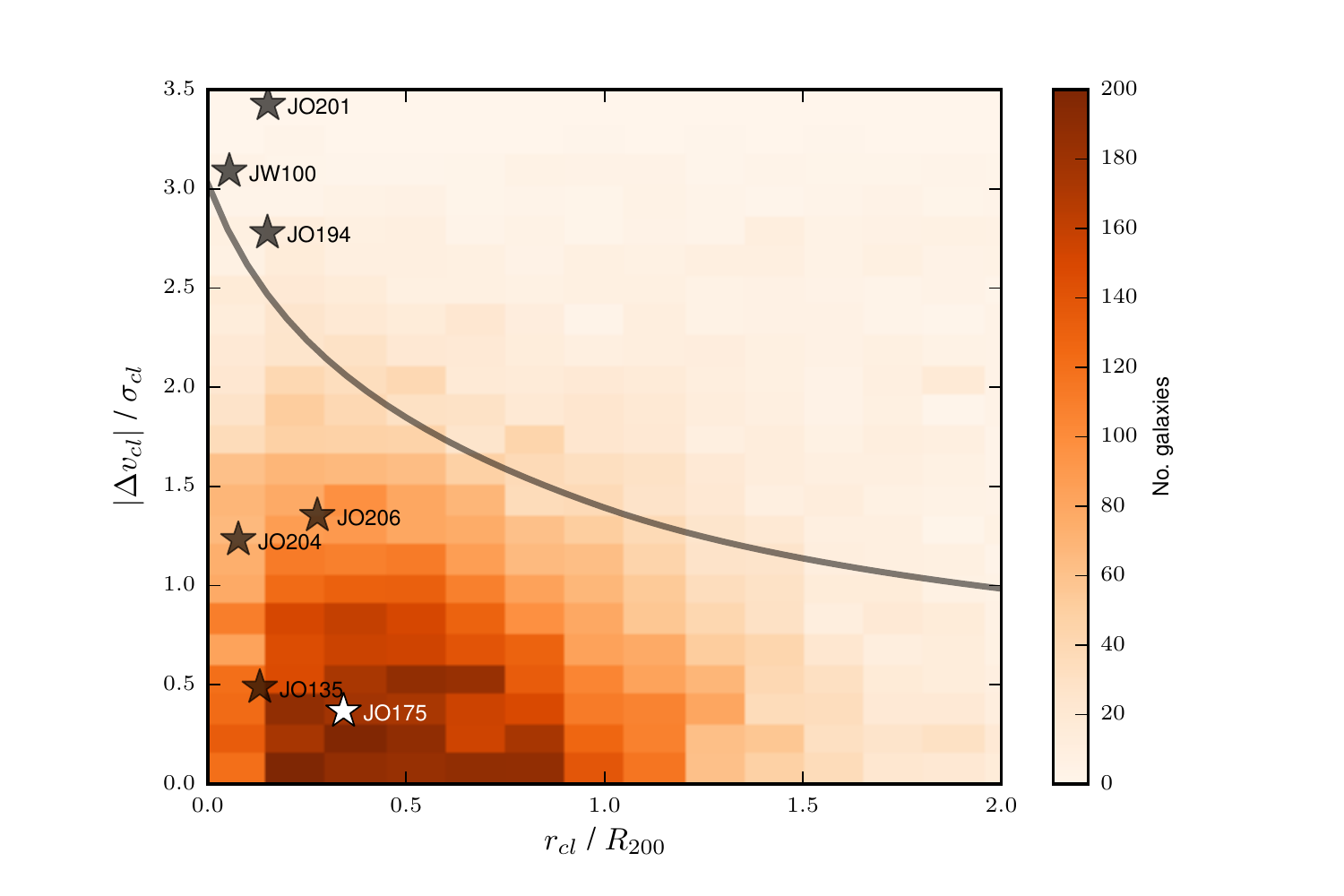}}
\caption{}
\end{figure*}

The main result is shown in Fig.~3.
We use standard diagnostic diagrams of emission-line ratios to 
assess the mechanism responsible for the gas ionization. 
The gas emitting in $\rm H\alpha$ can be ionized by different 
mechanisms: photons by young hot stars (Star-forming), the central 
AGN (AGN), a combination of the two (HII-AGN Composite) and Low 
Ionization Nuclear Emission-line Region  (LINER) that might be due to 
a low-luminosity AGN or other mechanisms such as shocks or old stars. 
To discriminate among Star-forming/HII-AGN Composite/LINER/AGN emission, we use the 
classification proposed by$^{18,19,20,21}$. 

According to the MUSE line ratios, the galaxy central regions are powered by
AGN emission in JO201, JO204, JW100, JO206 and JO135.  In JO194, the
central emission is LINER-like, as it is in a slightly larger annular
region surrounding a star-formation dominated ring. In contrast, line
ratios in JO175 are consistent with photoionization by star formation
in the center and throughout most of the disk and tails.

Thus, the great majority (5/7) of our jellyfishes host an AGN that is
evident from the MUSE spectra. This is at odds with the fact that
only 3\% of emission-line galaxies with a spectroscopic classification
in clusters at low redshift show evidence for AGN activity$^{22}$ (this
fraction is only slightly higher, $\sim 8\%$, among field galaxies$^{23}$). 
The AGN in our galaxies are responsible for the ionization in
a central region that is generally quite extended, up to 10kpc in
diameter (e.g. JO201, Fig.~3). 

Three of our galaxies (JO201, JO204, JW100) have two spectral
components with different velocities. The two  
components correspond to gas at different velocities that are seen in  
projection along the line of sight.
Interestingly, the two components in JO201 are both powered
by the AGN in the central region, while the two components of JO204
have a quite different spatial distribution: while the second
component is AGN-dominated in the central region, the first component
(JO204a) has an AGN-powered extraplanar region, extending up to 15kpc
away from the stellar isophotes, that appears to be an AGN-ionization
cone along the tails. Similarly, regions illuminated by the AGN are
seen out to large galactocentric distances in the disk of JO135, and
6kpc in projection outside of the stellar disk to the
north. Therefore, JO204 and JO135
possess a galaxy-scale ionization cone powered by the AGN.

The case of JO194 is more doubtful, as the LINER-like emission can be
due either to a low luminosity AGN or to other sources of ionization.
The spatial distribution of the LINER-like
emission favors the AGN origin.
Chandra (0.3-8keV) X-ray luminosities (Extended Data Table~1)
support our MUSE findings for AGN in
JO194 as well as in JO135, JW100, JO201 and JO206, the latter two being 
very X-ray luminous sources with $L_X=7.3 \times 10^{41} \, \rm erg \,
s^{-1}$ and $L_X=7.7 \times 10^{42} \, \rm erg \, s^{-1}$, respectively.
An independent proxy for the AGN luminosities are the [OIII]5007
luminosities, listed in Extended Data Table~1.
The conclusion that AGN emission is widespread in our jellyfish sample
is further reinforced in the summary diagram in Extended Data Figure~1.

The high incidence of AGN among the most striking jellyfish galaxies
uncovers a link between nuclear activity and strong ram pressure
stripping. Two scenarios can be envisaged. In the first one, the ram 
pressure is capable of funneling the gas towards the galaxy center,
causing gas accretion onto the central black hole and triggering the
activity. 
Hydrodynamic simulations have found that when galactic gas interacts 
with the non-rotating ICM it can lose angular momentum and spiral 
into the central region of a galaxy$^{24,25,26}$. 
Another possible method by which ram 
pressure stripping could feed an AGN is inflow of gas 
towards the galactic center generated by oblique shocks
in a disk that is flared due to the magnetic field $^{27}$.

The second scenario foresees the AGN injecting a large amount of
energy into the ISM, thus decreasing its binding energy and making it
more easily stripped, or even directly ejecting it from the galaxy$^{28}$.
In this case the AGN feedback would
increase the efficiency of ram pressure, and is an important component 
producing the striking jellyfish appearance.  

To discriminate between these two hypotheses, we show in Fig.~4 the
location of our jellyfishes in a projected position vs. velocity
phase-space diagram. 
The expected ram pressure increases with the ICM density, 
which gets higher going to the cluster center, and with the square of
the differential velocity$^{4}$. Thus, the most favorable conditions for ram pressure are 
at low radii and high $\Delta v_{cl}$$^{29}$, where most of our jellyfishes are
located (Fig.~4, see also Methods). 

Thus, the phase-space diagram strongly supports the hypothesis that it is ram
pressure that triggers the AGN, and not viceversa.  If the AGN were
making the ram pressure efficiency anomalously high, there is no reason
this should happen at the observed, most favorable location in the phase-space diagram.
This does not exclude that the energy injected by the AGN contributes
to an efficient gas loss, and helps creating the spectacular tails we
observe, with a sort of ``AGN-feedback'' in a cycle of ram pressure
triggering AGN favoring ram pressure. 
Simulations of ram pressure stripping including an AGN do not exist yet, but would be very
valuable for interpreting our discovery.

\section{Main references}
[1] Krawczynski, H. \& Treister E., Active galactic nuclei — the
physics of individual sources and the cosmic history of formation and
evolution, FrPhy, vol.8, issue 6, 609-629 (2013) \newline
[2] Heckman, T.M. \& Best, P.N., The Coevolution of Galaxies and
Supermassive Black Holes: Insights from Surveys of the Contemporary
Universe, Ann. Rev. Astron. Astrophys., 52, 589-660 (2014)\newline
[3] Boselli, A. \& Gavazzi, G., Environmental Effects on Late-Type Galaxies in Nearby Clusters, PASP 118, 517-559  (2006)\newline
[4] Gunn, J.E. \& Gott, J.R.,  On the Infall of Matter Into Clusters of Galaxies and Some Effects on Their Evolution, Astrophys. J., 176, 1-19  (1972)\newline
[5] Fumagalli, M. et al.,  MUSE sneaks a peek at extreme ram-pressure stripping events - I. A kinematic study of the archetypal galaxy ESO137-001, Mon. Not. R. Astron. Soc., 445, 4335-4344 (2014)\newline
[6] Ebeling, H. et al, Jellyfish: Evidence of Extreme Ram-pressure Stripping in Massive Galaxy Clusters, Astrophys. J., 781, L40-L44 (2014)\newline
[7] Magorrian, J., et al., The Demography of Massive Dark Objects in Galaxy Centers, Astron. J., 115, 2285-2305  (1998)\newline
[8] Gultekin, K. et al., The M-$\sigma$ and M-L Relations in Galactic Bulges, and Determinations of Their Intrinsic Scatter, Astrophys. J., 698, 198-221 (2009)\newline
[9] Sanders, D., et al., Ultraluminous infrared galaxies and the origin of quasars, Astrophys. J., 325, 74-91 (1988)\newline
[10] Hopkins, P. \& Hernquist, L., A Characteristic Division Between the Fueling of Quasars and Seyferts: Five Simple Tests, Astrophys. J., 694, 599-609 (2009) \newline
[11] Moore, B., et al., Galaxy harassment and the evolution of clusters of galaxies, Nature, 379, 613-616 (1996)\newline
[12] White, S.D.M., Rees, M.J., Core condensation in heavy halos - A two-stage theory for galaxy formation and clustering, Mon. Not. R. Astron. Soc., 183, 341-358 (1978) \newline
[13] Bekki, K. \& Couch, W., Starbursts from the Strong Compression of Galactic Molecular Clouds due to the High Pressure of the Intracluster Medium, Astrophys. J., 596, L13-L16 (2003)\newline
[14] Poggianti, B.M., et al., Jellyfish Galaxy Candidates at Low Redshift, Astron. J., 151, 78-97 (2016)\newline
[15] Poggianti, B.M., et al., GASP I: Gas stripping phenomena
  in galaxies with MUSE, Astrophys. J., in press arXiv 1704.05086 (2017) \newline
[16] Bellhouse, C., et al., GASP II. A MUSE view of extreme
  ram-pressure stripping along the line of sight: kinematics of the
  jellyfish galaxy JO201, Astrophys. J. in press arXiv 1704.05087 (2017)\newline
[17] Gullieuszik, M., et al., GASP IV: A muse view of extreme ram-pressure stripping in the plane of 
the sky: the case of jellyfish galaxy JO204, Astrophys. J. submitted (2017) \newline
[18] Kewley, L.J., et al., Optical Classification of Southern Warm Infrared Galaxies, Astrophys. J.S, 132, 37 (2001)\newline
[19] Kauffmann, G., et al.., The host galaxies of active galactic nuclei, Mon. Not. R. Astron. Soc., 346, 1055 (2003) \newline
[20] Sharp, R.G., Bland-Hawthorn, J., Three-Dimensional Integral Field Observations of 10 Galactic Winds. I. Extended Phase ($\geq$10 Myr) of Mass/Energy Injection Before the Wind Blows, Astrophys. J., 711, 818 (2010)\newline
[21] Kewley, L.J., et al., The host galaxies and classification of active galactic nuclei, Mon. Not. R. Astron. Soc., 372, 961-976 (2006)\newline
[22] Marziani, P., et al.,  Emission line galaxies and active galactic nuclei in WINGS clusters, Astron. Astrophys., 599, A83 (2017) \newline
[23] Brinchmann et al., The physical properties of star-forming galaxies in the low-redshift Universe, Mon. Not. R. Astron. Soc. 351, 1151-1179 (2004) \newline
[24] Schulz, S. \& Struck, C., Multi stage three-dimensional sweeping and annealing of disc galaxies in clusters, Mon. Not. R. Astron. Soc., 328, 185-202  (2001)\newline
[25] Tonnesen, S. \& Bryan, G.L., Gas Stripping in Simulated Galaxies with a Multiphase Interstellar Medium, Astrophys. J., 694, 789-804  (2009)\newline
[26] Tonnesen, S. \& Bryan, G.L., Star formation in ram pressure stripped galactic tails, Mon. Not. R. Astron. Soc., 422, 1609-1624 (2012)\newline
[27] Ramos-Martinez, M. \& Gomez, G.C., MHD simulations of ram 
  pressure stripping of disk galaxies, in Galaxies at high redshift and their evolution over cosmic time, IAU Symp. vol. 319, 143-143 (2016) \newline
[28] Bower, R., et al., Breaking the hierarchy of galaxy formation, Mon. Not. R. Astron. Soc., 370, 645-655 (2006)\newline
[29] Jaff\'e, Y. et al., BUDHIES II: a phase-space view of H I gas stripping and star formation quenching in cluster galaxies, Mon. Not. R. Astron. Soc., 448, 1715-1728 (2015) \newline
[30] Navarro, J.F., Frenk, C.S., White, S.D.M., A Universal Density Profile from Hierarchical Clustering, Astrophys. J., 490, 493-508 (1997)\newline

\section{Acknowledgements}
Based on observations collected at the European Organisation for Astronomical Research in the Southern Hemisphere 
under ESO programme 196.B-0578. 
We thank Matteo Fossati and Dave Wilman 
for developing and making available KUBEVIZ. We acknowledge financial 
support from PRIN-INAF 2014. B.V. acknowledges the support from 
an Australian Research Council Discovery Early Career Researcher Award 
(PD0028506). ST was supported by the Alvin E. Nashman Fellowship in Theoretical Astrophysics.
This work was co-funded under the Marie Curie Actions of the European Commission (FP7-COFUND).

\section{Authors contribution} 
All authors contributed to the interpretation of the observations and
the writing of the paper.
B.M.P. led the project and performed the data analysis. 
Y.J. performed the phase-space analysis. A.M. carried
out the stellar kinematics analysis. M.G. did the data
reduction. M.R. contributed to the data analysis.
S.T. provided the discussion on simulations.  J.F. did the SINOPSIS analysis.
D.B. and G.F. helped in the preparation of the observations.
B.V. performed a comparison of the stellar population analysis 
and prepared the GASP web page. 
C.B. performed the two component KUBEVIZ analysis of JO201.
G.H. did the data reduction for JO201.
A.O. selected the JW100 target.

\section{Authors information} 
Reprints and permissions information is available at
www.nature.com/reprints.

The authors declare no competing financial interests.

Correspondence and requests for materials should be addressed to bianca.poggianti@oapd.inaf.it.

\section{Main figure legends}

{\bf Fig.~1} 
{\bf TITLE: MUSE stellar velocity map and $\rm H\alpha$ map for JO201, JO204 and JW100.}
MUSE stellar velocity map ((a) panels) and $\rm H \alpha$
velocity map ((b) and (c) panels) 
of our jellyfish galaxies. JO201, JO204 and JW100 have regions with two line components separated 
  in velocity, and their gas velocity maps are plotted separately
  (panels b) and c)). 
 Contours in all panels are stellar isophotes and indicate where the 
 galaxy stellar disk is. 
In the a) panels, the scale in kpc is indicated by a 
  bar and the arrow points in the direction of the cluster center. 
North is up and east is left.

{\bf Fig.~2} 
{\bf TITLE: MUSE stellar velocity map and $\rm H\alpha$ map for JO206, JO135, JO194 and JO175.}
As Fig.~1 for the other 4 galaxies.

{\bf Fig.~3} 
{\bf TITLE: Diagnostic diagrams and maps for all jellyfishes.}
Spatially resolved diagnostic diagrams ((a) panels) and
maps ((b) panels) for 
  all MUSE pixels where lines are measured with a signal-to-noise$>3$.
For JO201, JO204 and JW100 the two components are presented separately
and there are 4 panels per galaxy. In (a) panels, lines$^{18,19,20}$ separate
Star-forming, HII-AGN Composite, AGN and LINERS.  Only in the case of
JW100, lines$^{21}$ separate Star forming, AGN and LINERs. Contours
are stellar isophotes, as in Fig.~1.  For each galaxy we have
inspected both the [OIII]5007/$\rm H\beta$ vs. [NII]6583/$\rm H\alpha$
and the [OIII]/$\rm H\beta$ vs. [SII]6717/$\rm H\alpha$ diagrams and found no discrepancy 
of classification between the two. For convenience, we show only the 
spatially resolved [NII]6583/$\rm H\alpha$ plot for each galaxy, except for JW100 for which 
we use the [SII]6717/$\rm H\alpha$ plot instead, because at the JW100 
redshift the [NII] line is contaminated by a sky line. 

{\bf Fig.~4}
{\bf TITLE: Differential velocity versus clustercentric distance.}
Phase-space diagram: projected differential velocity with 
  respect to the cluster  median velocity, normalized by the cluster velocity dispersion,
  versus the projected clustercentric distance, in units of cluster 
  virial radius $R_{200}$. The latter is defined as the projected radius delimiting  
a sphere with interior mean density 200 times the critical density of  
the Universe. In this plot, velocities and radii are lower limits to  
the three dimensional velocity of the galaxy through the ICM and  
clustercentric distance, respectively.  
The location of our jellyfishes is 
  signposted by the stars. The only jellyfish with no AGN, JO175, is 
  marked with a white star. The number density of all cluster 
  galaxies from the OMEGAWINGS sample at each 
  location in the diagram is color coded (see bar on the right hand 
  side). The darker orange regions trace the location of the oldest 
  cluster members, that live near the cluster core (at low 
$|\Delta v_{cl}|/\sigma_{cl}$) after having settled into the potential 
well. Thus, the position of the jellyfish galaxies in phase-space implies that they are 
being stripped on first infall onto che cluster. 
The curve represents the escape velocity in 
  a dark matter 
halo$^{30}$. 

\section{Methods}
In this work we adopt a standard concordance cosmology with 
$H_0 = 70 \, \rm km \, s^{-1} \, Mpc^{-1}$, ${\Omega}_M=0.3$
and ${\Omega}_{\Lambda}=0.7$ and a stellar Initial Mass Function from$^{31}$. 
The OMEGAWINGS spectroscopic catalog used to generate Fig.~4 
is taken from$^{32}$.

\subsection{Observations and line fitting}
The galaxies analyzed in this paper have been observed by the 
GASP program with 1 or 2 (depending on the lenght of the tails) MUSE 
pointings of 2700sec each in service mode, with seeing conditions 
$\leq$ 1 arcsec. The MUSE spectrograph$^{33}$ has a 1'X1' 
field-of-view with 0.2''X0.2'' 
pixels with a spectral range 4800-9300\AA $\,$ at 2.6\AA $\,$ resolution. Prior 
to the analysis, the datacube is average-filtered in the spatial 
dimension with a 5X5 pixel kernel, corresponding to 1 arcsec (the 
upper limit of the seeing)=0.8-1.1 kpc 
depending on the galaxy redshift. No smoothing nor binning is 
performed in the spectral direction. 
The observations, data reduction and analysis tools are described in 
details in $^{15}$ . 

Emission lines in the datacube are fitted with gaussian profiles with 
KUBEVIZ$^{34}$ , a public IDL software that uses the MPfit package and 
provides gas velocities (with respect to a given redshift), velocity 
dispersions and line fluxes. KUBEVIZ can attempt a single or a double 
component fit (see $^{16}$  for details). 
Three of the galaxies presented in this work -- JO201, JO204 and JW100 -- require  a 
double component fit, for which we have shown velocity and diagnostic 
diagrams for each one of the two components separately. 
None of these galaxies have a broad 
component in permitted lines (Seyfert1), with $\rm H\alpha$ widths ($\sigma$) up to a 
few hundreds km per sec. 

In JO135, there is a small central region (white in Fig.~3)   
where a line gaussian (even double) fit cannot be obtained. 
Inspecting the MUSE spectra,  it is clear that this is due to the   
very strong asymmetry of the lines indicating 
a very powerful nuclear outflow. We note that the literature 
reports an 8kpc AGN outflow in another jellyfish galaxy, NGC 4569 in 
the Virgo cluster$^{35}$ . 

The line intensities in Extended Data Figure~1 are measured from KUBEVIZ in 
mask mode, masking out all the spaxels outside of the region of 
interest. In this case KUBEVIZ was run in interactive mode, to verify visually the quality of the fit. 
The errorbars are computed propagating the KUBEVIZ errors on the 
line fluxes, and are small thanks to the very high signal-to-noise of
the spectra. 

\subsection{Analysis techniques}
The results shown in Fig.~3 have been obtained from the datacube corrected both 
for Galactic extinction and for intrinsic dust extinction calculated from 
the $\rm H\alpha$/$\rm H\beta$ ratio$^{15}$ and after having subtracted the stellar 
component using the spectrophotometric fits of the code SINOPSIS$^{36}$.  
This code, fully described in $^{36}$, searches the combination of single stellar population 
(SSPs) spectra that best fits the observed equivalent widths of the main lines in absorption 
and in emission and the continuum at various wavelengths,
minimizing the ${\chi}^2$ using an Adaptive Simulated Annealing 
algorithm. 
The current version of SINOPSIS uses the latest SSPs model from Charlot \& Bruzual (in 
prep.) that have a higher spectral and age resolution than previous versions 
and cover metallicity values from $Z=0.0001$ to $Z=0.04$. These 
models use the latest evolutionary tracks from$^{37}$ 
and stellar atmosphere emission from a compilation of different authors. 
Moreover, SINOPSIS includes nebular emission 
for the youngest (i.e.age $< 2\times 10^7$ years) SSP, computed 
ingesting the original models into the plasma simulation code 
CLOUDY$^{38}$. 
SINOPSIS provides spatially resolved maps of stellar masses, star 
formation rates, star formation histories, luminosity-weighted ages 
and other stellar population properties. 
The total galaxy stellar masses listed in Extended Data Table~1 are computed summing up the stellar 
mass in each spaxel estimated from SINOPSIS. 

The stellar kinematics is derived using the Penalized Pixel-Fitting 
code$^{39}$, with the method presented in $^{15}$. 
This code fits the observed spectra with the stellar population 
templates by$^{40}$, 
using SSPs of 6 different metallicities (from $[M/H]=-1.71$ to $[M/H]=0.22$) 
and 26 ages, from 1 to 17.78 Gyr. 
After having accurately masked spurious sources (stars, 
background galaxies) in the galaxy proximity, 
and having degraded the spectral library resolution to our MUSE 
resolution, 
we performed the fit of spatially binned spectra based 
on signal-to-noise (S/N=10, for most galaxies), as described 
in$^{41}$, with the Weighted Voronoi Tessellation 
modification proposed by$^{42}$. 
This yields maps of the rotational velocity, the velocity dispersion and the two h3 and 
h4 moments using an additive Legendre polynomial fit of the 12th order 
to correct the template continuum shape during the fit. 

The gas velocity map (Figs.~1 and ~2) is obtained from the absorption corrected cube average filtering in the spatial directions with a 
  5$\times$5 pixel kernel, plotting only spaxels with a ${S/N}_{\rm 
    H\alpha}>4$. The stellar map is shown for the Voronoi bins 
 with a $S/N>10$. We note that in Figs.~1 and ~2 the gaseous and stellar 
 velocity zero points are coincident and correspond to the galaxy 
 redshift listed in Extended Data Table~1, 
  except for JW100 where the stellar zeropoint is at redshift z=0.06214 because 
  gas and stars have a large systematic shift. 
The contours in Figs.~1 and ~2 are logarithmically spaced isophotes of the 
 spectral continuum underlying $\rm  
  H\alpha$, thus are stellar isophotes, down to a surface brightness  
  $2.5 \times 10^{-18} \rm erg \, s^{-1} \, cm^{-1} \, {\AA}^{-1} \,
  arcsec^{-2}$. 

As mentioned above, LINER-like emission-line ratios (above the
  solid line and to the right of the dashed line in Extended Data Figure~1)
can originate from a variety of physical processes$^{43, 44, 45, 46}$.
In constrast, the Seyfert-like line ratios (above the solid line and
to the left of the dashed line in Extended Data Figure~1) of
  JO201, JO204, JW100, JO206 and JO135 identify these galaxies as AGN.
This conclusion is further strengthened by the
equivalent widths of $\rm H\alpha$ and [OIII]5007 measured from
the integrated spectra of the region powered by the AGN, whose
rest-frame, absorption-corrected values, given in Extended Data Table~1, 
are higher than the low values measured in LINERs$^{46}$, typically
$EW(\rm H\alpha)< 3$ \AA.

Shocks induced by gas flows (in our case, by ram pressure) can give rise to line ratios that occupy also 
the ``AGN'' locus in the diagnostic diagrams$^{47}$,  
however the spatial distribution of the AGN-dominated 
spaxels, at the galaxy center, makes it very unlikely this is due to 
ram pressure shocks (which would be observed at the shock fronts with 
the ICM), and strongly favors the AGN hypothesis. 

The ram pressure can be computed$^{4}$ as 
$P_{ram} = \rho_{ICM} \times \Delta v^2_{cl}$, 
where $ \rho_{ICM} $ is the ICM density and $\Delta v^2_{cl}$ is the 
differential galaxy velocity with respect to the cluster, as in Fig.~4. 
Figure~4 shows that most of our jellyfishes are indeed in the conditions of 
strong ram pressure, being at very high (JO204,JO206,
$|\Delta v_{cl}|/\sigma_{cl} > 1$) or extremely high (JO201, JW100 
and JO194, $|\Delta v_{cl}|/\sigma_{cl} > 2.5$) velocities, and 
very small (projected) radii. 
JO135 is at a small projected clustercentric radius, but its relative 
radial velocity is lower than the other AGN. However, 
its 3D velocity relative to the ICM might be much larger if the tangential 
velocity (along the plane of the sky) is much higher than the radial 
velocity, as suggested by Fig.~2. Moreover, JO135 is part of the 
Shapley supercluster and it is located at a position where the two 
clusters A3532 and A3530 are merging, and this likely causes a ram 
pressure enhancement$^{48}$. 
Interestingly,  JO175, that is the only 
jellyfish with no evidence for an AGN, lies at low relative radial 
velocity $|\Delta v_{cl}|/\sigma_{cl} \sim 0.3$. 

\subsection{Code availability}
This work made use of the KUBEVIZ software which is publicly available
at \newline 
http://www.mpe.mpg.de/$\sim$dwilman/kubeviz/, 
of the Voronoi binning and pPXF software available at
 http://www-astro.physics.ox.ac.uk/$\sim$mxc/software/,
and the SINOPSIS code 
that is publicly available under the MIT open source licence and can 
be downloaded from \newline
http://www.crya.unam.mx/gente/j.fritz/JFhp/SINOPSIS.html. 

\subsection{Data Availability Statement}
The MUSE data that support the findings of this study are part of the
Phase3 data release of the GASP program and will be
available in the ESO Archive at http://archive.eso.org/cms.html. The
first GASP public data release, including the data
regarding this article, will be released at the end of 2017. 

\section{Additional references used in the methods}
[31] Chabrier, G., Galactic Stellar and Substellar Initial Mass Function, PASP, 115, 763-795 (2003) \newline
[32] Moretti, A. et al., OmegaWINGS: spectroscopy in the outskirts of local clusters of galaxies, Astron. Astrophys., 599, A81 (2017) \newline
[33] Bacon, R., et al., The MUSE second-generation VLT instrument,
SPIE, 7735, id. 773508 (2010) \newline
[34] Fossati, M., et al., MUSE sneaks a peek at extreme ram-pressure stripping events - II. The physical properties of the gas tail of ESO137-001, Mon. Not. R. Astron. Soc., 455, 2028-2041 (2016) \newline
[35] Boselli, A., et al., Spectacular tails of ionized gas in the Virgo cluster galaxy NGC 4569, Astron. Astrophys., 587, A68 (2016) \newline
[36] Fritz, J., et al., GASP III. JO36: a case of multiple environmental effects at play?, Astrophys. J. submitted arXiv 1704.05088 (2017) \newline
[37] Bressan, A., et al., PARSEC: stellar tracks and isochrones with the PAdova and TRieste Stellar Evolution Code, Mon. Not. R. Astron. Soc., 427, 127-145 (2012) \newline
[38] Ferland, G.J., et al., The 2013 Release of Cloudy, MxAA, 49, 137-163 (2013) \newline
[39] Cappellari, M., Emsellem, E., Parametric Recovery of Line-of-Sight Velocity Distributions from Absorption-Line Spectra of Galaxies via Penalized Likelihood, PASP, 116, 138-147 (2004) \newline
[40] Vazdekis, A., et al., Evolutionary stellar population synthesis with MILES - I. The base models and a new line index system, Mon. Not. R. Astron. Soc., 404, 1639-1671 (2010) \newline
[41] Cappellari, M., Copin, Y., Adaptive spatial binning of integral-field spectroscopic data using Voronoi tessellations, Mon. Not. R. Astron. Soc., 342, 345-354  (2003) \newline
[42] Diehl, S., Statler, T.S., Adaptive binning of X-ray data with weighted Voronoi tessellations, Mon. Not. R. Astron. Soc., 368, 497-510 (2006) \newline
[43] Sarzi, M., et al., The SAURON project - XVI. On the sources of
ionization for the gas in elliptical and lenticular galaxies, Mon. Not. R. Astron. Soc., 402, 2187-2210 (2010)\newline
[44] Yan, R. \& Blanton, M.R., The Nature of LINER-like Emission in
Red Galaxies, Astrophys. J., 747, id. 61 (2012)\newline
[45] Singh, R., et al., The nature of LINER galaxies:. Ubiquitous hot
old stars and rare accreting black holes, Astron. Astrophys., 558, A43 (2013) \newline
[46] Belfiore, F., et al., SDSS IV MaNGA - spatially resolved
diagnostic diagrams: a proof that many galaxies are LIERs, Mon. Not. R. Astron. Soc., 461,
3111-3134 (2016) \newline
[47] Allen, M.G., et al., The MAPPINGS III Library of Fast Radiative
  Shock Models, Astrophys. J.S, 178, id. 20-55 (2008) \newline
[48] Owers, M. et al.,  Shocking Tails in the Major Merger Abell 2744,
Astrophys. J., 750, id. L23 (2012) \newline
[49] Cava, A. et al., WINGS-SPE Spectroscopy in the WIde-field Nearby Galaxy-cluster Survey, Astron. Astrophys., 495, 707-719 (2009) \newline
[50] Wang, S. et al., CHANDRA ACIS Survey of X-Ray Point Sources: The
Source Catalog, Astrophys. J.S, 224, id. 40 (2016) \newline

\section{Extended Data table legend} 

{\bf Title: Properties of GASP jellyfish galaxies.}
The IDs of our galaxies (as given by  $^{14}$), their
host cluster name, cluster 
velocity dispersion$^{30,49}$, galaxy coordinates,
redshifts, stellar masses, X-ray luminosities (from $^{50}$), GASP
[OIII]5007 luminosities  and rest frame emission-only equivalent widths (EWs) of $\rm
H\alpha$ and [OIII]5007
are listed in Extended Data Table~1. [OIII]5007 luminosities and EWs
have been computed on the 
absorption-corrected integrated spectra of the AGN regions
(LINER for JO194, and central star-forming region for JO175), see the
caption of Extended Data Figure~1. 
In case of galaxies with two components, the
[OIII] luminosity is the sum of the two luminosities and the two EWs
are listed separated by a slash. The sum of these two EWs can be
thought of as a ``total'' EW. Other properties of these 
galaxies (gas and stellar kinematics, stellar history, gas metallicity
and others) are the subject of dedicated publications$^{15,16,17}$.

\section{Extended Data figure legend} 

{\bf Title: Summary diagnostic diagrams}
{\bf Extended Data Figure~1} Line ratio diagrams summarizing our findings showing the location of each galaxy in two different diagnostics diagrams 
integrating the spectrum over the spatial region (identified from 
Fig.~3) dominated by AGN emission (JO201, JO204, JW100, JO206, JO135),
by LINER emission (JO194) and over the central 7$\times$7 brightest spaxels in 
the case of JO175. Here we present both the [NII]6583/$\rm H\alpha$ and  
the [SII]6717/$\rm H\alpha$ diagrams, to illustrate the good agreement  
between the two and to display also JW100 whose [NII] line cannot be  
measured. 
Lines 
as in Fig.~3. 
The two 
components in JO201, JO204 and JW100 are shown as separate points. 
The errorbars are computed propagating the errors on the line fluxes
obtained by KUBEVIZ, scaled to achieve a reduced $\chi^2=1$ as
described in$^{15}$. 

\begin{figure*}
\centerline{\includegraphics[scale=1.0]{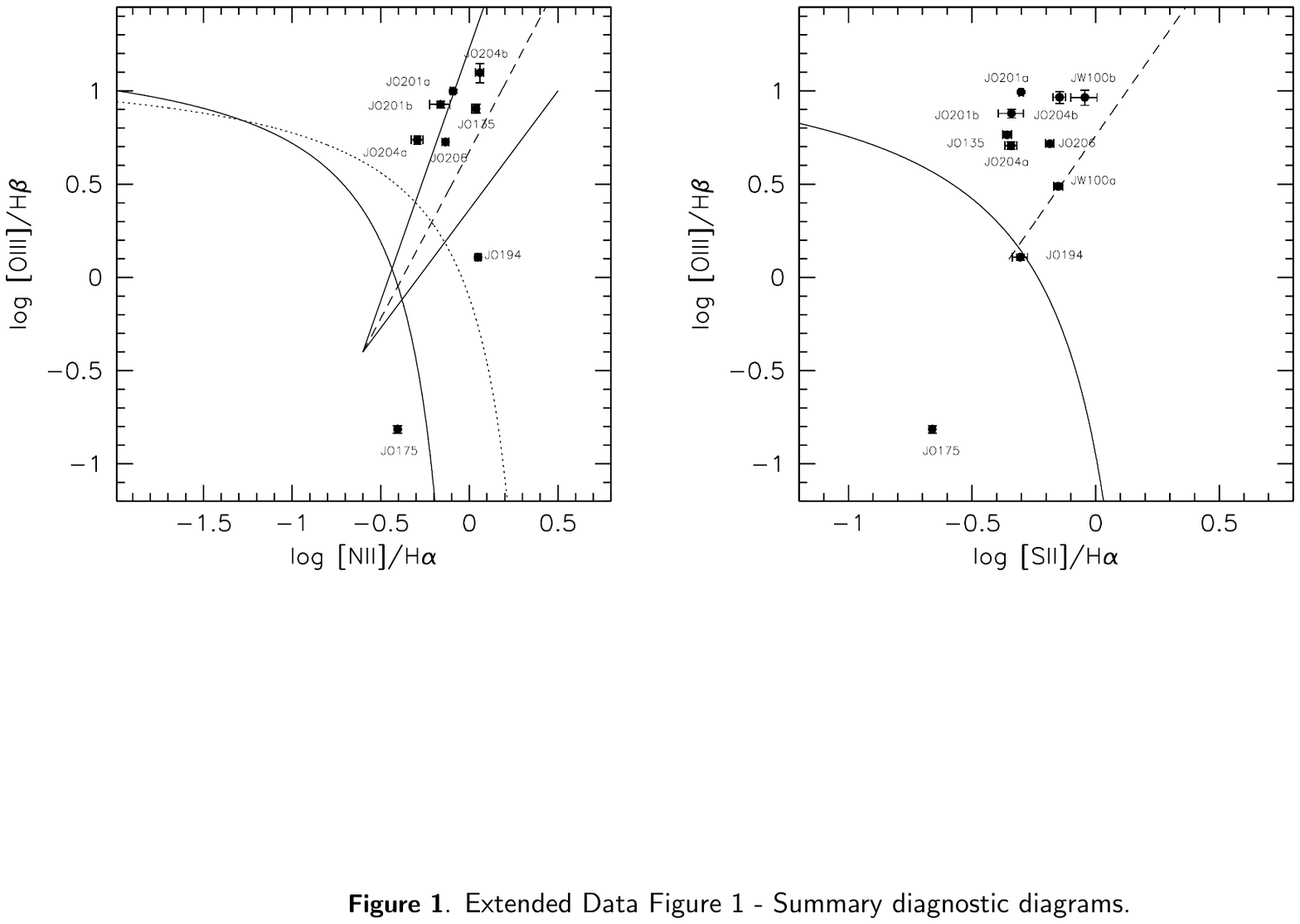}}
\end{figure*}

\begin{figure*}
\centerline{\includegraphics[scale=1.0]{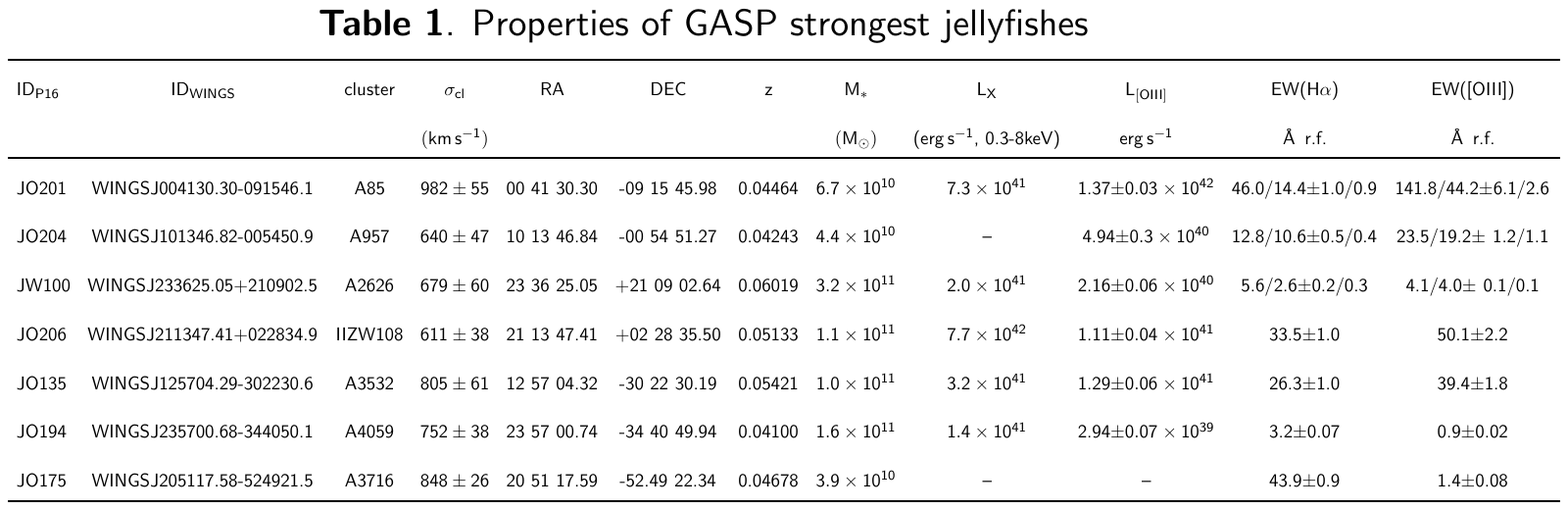}}
\end{figure*}

\end{document}